\documentclass[showpacs,prb,twocolumn,floats]{revtex4}

\usepackage{graphicx}
\usepackage{amsmath}
\usepackage{amssymb}
\usepackage[colorlinks=true,citecolor=blue,linkcolor=blue]{hyperref}
\usepackage{amsfonts}
\usepackage[latin9]{inputenc}
\usepackage{color}
\usepackage{comment}
\usepackage{epstopdf}

\renewcommand{\vec}[1]{\ensuremath{\boldsymbol{#1}}}

\begin{document}

\title{Transport properties of bilayer graphene in a strong in-plane magnetic field}
\date{\today }
\author{M. Van der Donck}
\email{matthias.vanderdonck@uantwerpen.be}
\affiliation{Department of Physics, University of Antwerp, Groenenborgerlaan 171, B-2020 Antwerp, Belgium}
\author{F. M. Peeters}
\email{francois.peeters@uantwerpen.be}
\affiliation{Department of Physics, University of Antwerp, Groenenborgerlaan 171, B-2020 Antwerp, Belgium}
\author{B. Van Duppen}
\email{ben.vanduppen@uantwerpen.be}
\affiliation{Department of Physics, University of Antwerp, Groenenborgerlaan 171, B-2020 Antwerp, Belgium}


\begin{abstract}
A strong in-plane magnetic field drastically alters the low-energy spectrum of bilayer graphene by separating the parabolic energy dispersion into two linear Dirac cones. The effect of this dramatic change on the transport properties strongly depends on the orientation of the in-plane magnetic field with respect to the propagation direction of the charge carriers and the angle at which they impinge on the electrostatic potentials. For magnetic fields oriented parallel to the potential boundaries an additional propagating mode that results from the splitting into Dirac cones enhances the transmission probability for charge carriers tunneling through the potentials and increases the corresponding conductance. Our results show that the chiral suppression of transmission at normal incidence, reminiscent of bilayer graphene's $2\pi$ Berry phase, is turned into a chiral enhancement when the magnetic field increases, thus indicating a transition from a bilayer to a monolayer-like system at normal incidence. Further, we find that the typical transmission resonances stemming from confinement in a potential barrier are shifted to higher energy and are eventually transformed into anti-resonances with increasing magnetic field. For magnetic fields oriented perpendicular to the potential boundaries we find a very pronounced transition from a bilayer system to two separated monolayer-like systems with Klein tunneling emerging at certain incident angles symmetric around 0, which also leaves a signature in the conductance. For both orientations of the magnetic field, the transmission probability is still correctly described by pseudospin conservation. Finally, to motivate the large in-plane magnetic field, we show that its energy spectrum can be mimicked by specific lattice deformations such as a relative shift of one of the layers. With this equivalence we introduce the notion of an in-plane pseudo-magnetic field.
\end{abstract}

\maketitle

\section{Introduction}

Since the experimental realisation of graphene,\cite{ontdekking} a lot of research has been done in order to understand its remarkable electronic structure and transport properties. Notably, charge carriers in graphene were shown, both theoretically\cite{klein} and experimentally,\cite{kleinexp,kleinexp2} to behave like ultra-relativistic particles which lead to the introduction of the concept of pseudospin and the observation of Klein tunneling.\cite{klein} It is also remarkable that when multiple graphene sheets are stacked the low-energy electronic structure completely changes, despite the very weak van der Waals coupling between the graphene sheets. This is a consequence of the fact that the stacked multilayer structures still have a high symmetry and results in very different transport properties such as anti-Klein tunneling for bilayer graphene\cite{vierband, Tudorovskiy2012, Snyman2007} and even more remarkable phenomena for multilayers.\cite{multiband,multilaag}

An in-plane magnetic field has no influence on the energy-momentum relation of monolayer graphene but it does alter the electronic properties of multilayer and specifically bilayer graphene.\cite{pseudo, InPlaneField2} In contrast to a perpendicular magnetic field, for which the influence on the transport properties of both monolayer\cite{monoloodrecht} and bilayer\cite{biloodrecht} graphene has been investigated in much detail, an in-plane magnetic field breaks the equivalence of the two layers of bilayer graphene and as such decouples them at low energy.

In this work we investigate how an in-plane magnetic field influences the electronic and transport properties of electrons in bilayer graphene that tunnel through potential steps and barriers. We show that the results can be explained by using the chiral properties of the charge carriers.\cite{klein} The crucial effect of an in-plane magnetic field is that it changes the parabolic low-energy spectrum of bilayer graphene into two linear Dirac cones each corresponding to one of the two graphene layers.\cite{InPlaneField1, InPlaneField2} For magnetic fields oriented parallel to the potential boundaries this leads to an additional propagating mode that strongly affects the tunneling properties by increasing the conductance and introducing Klein tunneling at normal incidence and at strong enough magnetic fields, reminiscent of the linear electron dispersion. Further, we find that the transmission resonances that are due to confinement in the barrier are morphed into anti-resonances at large magnetic fields. These features are visible in the conductance of the system and in this way we provide a relation between empirically verifiable quantities as the conductance,\cite{condexp,condexp2} Fermi level and the magnetic field. For magnetic fields oriented perpendicular to the potential boundaries the splitting into Dirac cones leads to a transition from a bilayer system to two separated monolayer-like systems with Klein tunneling emerging at certain incident angles symmetric around 0, a feature which can also be seen in the conductance.

The strength of the magnetic field used in this paper is very large and although these fields can not be created in terrestrial laboratories, we show that they can be achieved by means of shifting the two graphene planes with respect to each other. Following a similar reasoning as for monolayer graphene\cite{Guinea2010} in this way one creates an in-plane pseudo-magnetic field.

We are aware of two articles which study similar systems. However, the first paper\cite{pseudo} only focuses on one orientation of the magnetic field and the second paper\cite{zelfde2} only focuses on the case in which the Fermi energy equals the height of the potential barrier. Furthermore, both papers were limited to transport governed by a single transmission channel. Our findings match these results in the respective limits but we extend them by taking into account all the different transmission and reflection channels for a range of both Fermi energy and incident angle values and for two fundamentally different orientations of the in-plane magnetic field. This is important since it shows new results. Furthermore, we calculate the conductance which also is a new and important result because it shows the experimentally observable signatures of the new physics.

The paper is organised as follows. In Sec. \ref{sec:Energy spectrum and pseudospin} we discuss the influence of an in-plane magnetic field on the energy spectrum of bilayer graphene and on the chiral properties of the charge carriers. We discuss how very strong pseudo-magnetic fields can be created in shifted bilayer graphene in Sec. \ref{sec:Creating pseudo-magnetic fields using shifted bilayer graphene}. In Sec. \ref{sec:Spinor wave function} we calculate the spinor wave function from which the current density is determined in Sec. \ref{sec:Current density and scattering probabilities} which is needed to calculate the different transmission and reflection probabilities. The numerical results of the different scattering probabilities are discussed in Sec. \ref{subsec:Scattering probabilities} and with them the conductance is determined and discussed in Sec. \ref{subsec:Conductance}. Finally, in Sec. \ref{sec:Conclusion} we summarize the main conclusions of this paper.

\section{Energy spectrum and pseudospin}
\label{sec:Energy spectrum and pseudospin}

\begin{figure}[tb]
\centering
\includegraphics[width= 8.5cm]{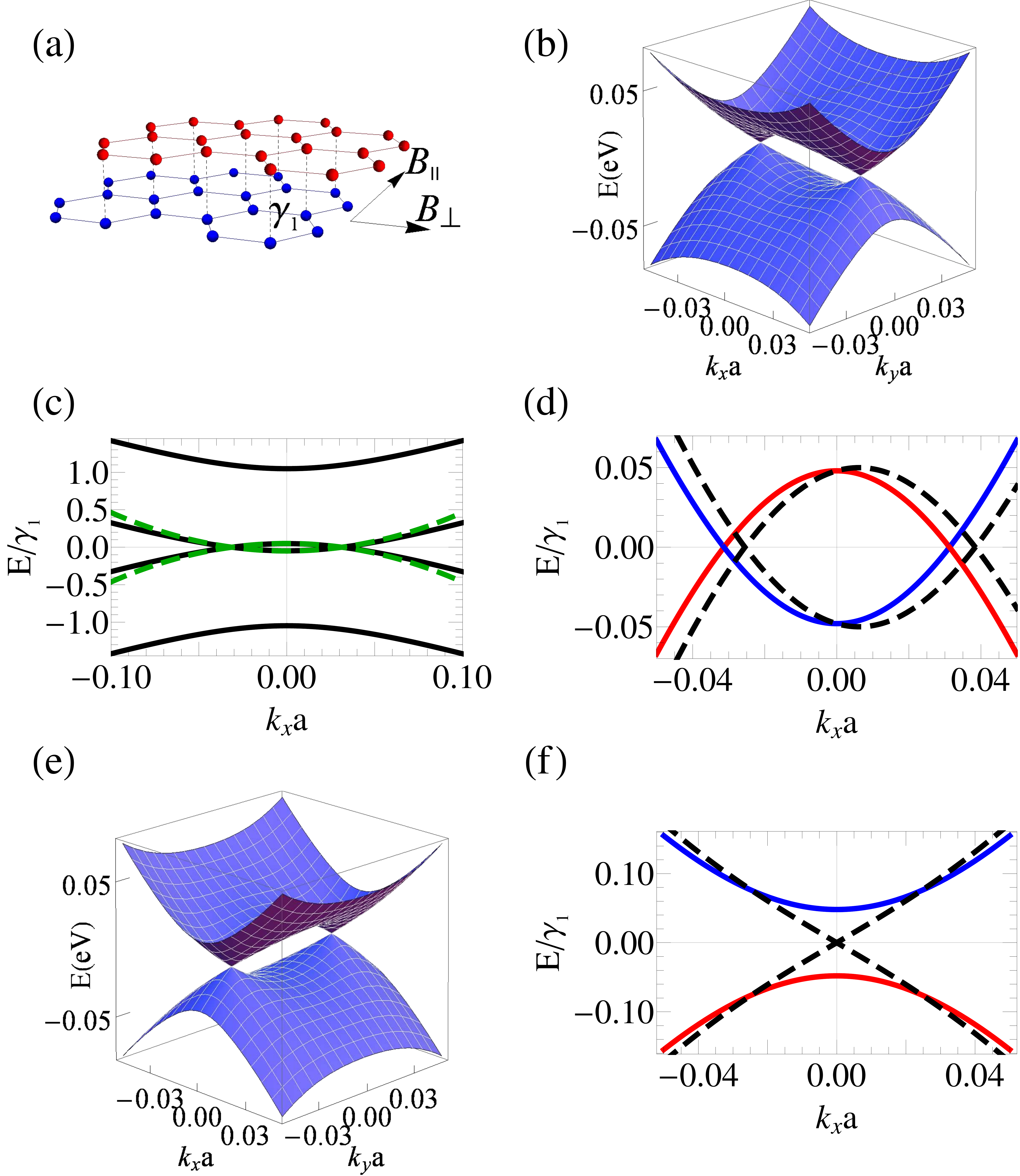}
\caption{(Color online) (a) Illustration of Bernal stacked bilayer graphene with the indication of the two magnetic field orientations that are considered in this work and the interlayer hopping parameter. (b) 3D plot of the low-energy bands in the presence of an in-plane magnetic field oriented parallel to the potential boundaries with strength $B=500$ T showing the presence of the two low-energy cones. (c) Slice of the four band energy spectrum in (b) (black solid curves) and two band energy spectrum (green dashed curves) for $k_y=0$. (d) Zoom of the two low-energy bands in (c). The red energy band corresponds with the pseudospin-up state, the blue energy band is the pseudospin-down state. The dashed black energy bands include the $\gamma_3$ hopping parameter. (e) Same as (b) for a magnetic field oriented perpendicular to the potential boundaries. (f) Slice of the energy spectrum in (e) for $k_y=0$ (red curve for the pseudospin-up state and blue curve for the pseudospin-down state) and for $k_y=\kappa$ (black dashed curve).}
\label{fig:energie}
\end{figure}

Bilayer graphene consists of two stacked hexagonal monolayers which in turn are made of two trigonal sublattices. We denote them as $A$ and $B$ for the bottom layer and $A'$ and $B'$ for the top layer. The intralayer interatomic distance is $a_0=1.42$ \AA, which is related to the lattice constant $a=2.46$ \AA\ by $a=\sqrt{3}a_0$ and the interlayer distance measures $c=3.35$ \AA. In this paper we consider Bernal stacked bilayer graphene\cite{bernal} for which the $B$ and $B'$ sublattices are situated above each other as illustrated in Fig. \ref{fig:energie}(a). The coupling between these sublattices is therefore the dominant interlayer interaction and has a corresponding hopping parameter $\gamma_1=0.377$ eV in the tight-binding approximation.\cite{parameters} The intralayer hopping parameter is $\gamma_0=3.12$ eV.\cite{parameters}

\begin{figure}[tb]
\centering
\includegraphics[width= 8.5cm]{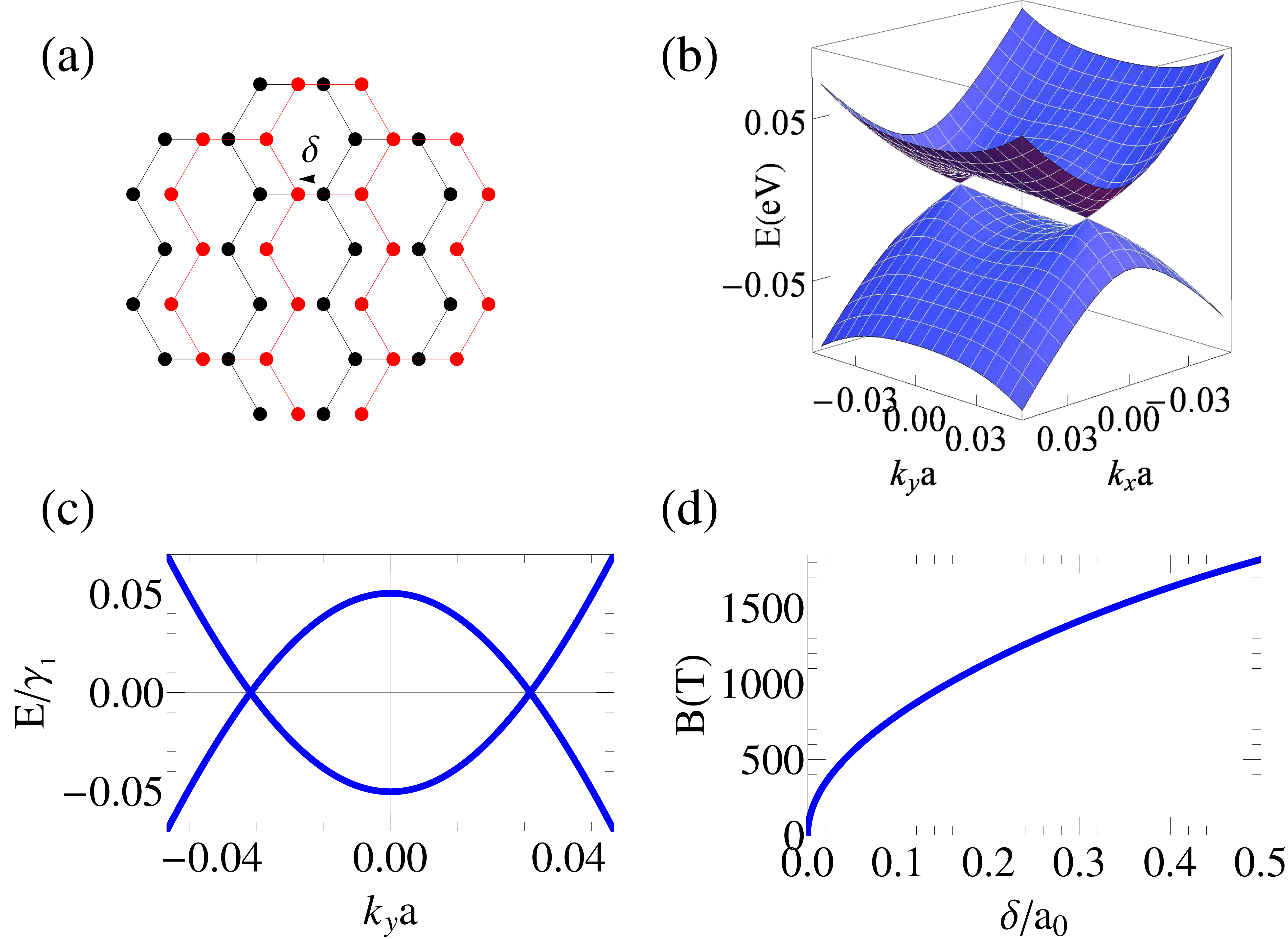}
\caption{(Color online) (a) Illustration of the shifted bilayer graphene lattice. The bottom and top layer are shown in black and red, respectively. (b) 3D plot of the low-energy bands of shifted bilayer graphene with $\delta=0.04a_0$. (c) Slice of the low-energy spectrum of shifted bilayer graphene for $k_x=0$. (d) Pseudo-magnetic field as a function of the shift of the top layer.}
\label{fig:shiftplot}
\end{figure}

The presence of a magnetic field parallel to the graphene plane can be incorporated by means of the Peierls substitution $\vec{k}\rightarrow \vec{k}+\frac{e}{\hbar}\vec{A}$.\cite{InPlaneField1, InPlaneField2} Two possible choices of the gauge field are considered: $\vec{A}_{\parallel}(z)=Bz\vec{e}_x$ corresponding to a magnetic field $\vec{B}_{\parallel}=B\vec{e}_y$ and $\vec{A}_{\perp}(z)=-Bz\vec{e}_y$ corresponding to a magnetic field $\vec{B}_{\perp}=B\vec{e}_x$, where $\vec{e}_x$ and $\vec{e}_y$ are the unit vector in the $x$- and $y$-direction, respectively. These two cases are indicated in Fig. \ref{fig:energie}(a) and will have very different effects on the transport properties as the magnetic fields are respectively oriented parallel and perpendicular to the boundaries of the potentials that we will consider. If we choose $z=-c/2$ for the bottom layer and $z=c/2$ for the top layer, the Hamiltonian $H$ becomes in the basis of the atomic Bloch functions\cite{tightbinding} $\Psi = (\psi_A,\psi_B,\psi_{A'},\psi_{B'})^T$ 
\begin{equation}
\label{hamiltoniaan}
H_{\alpha} = \hbar v_{\rm F}
\begin{pmatrix}
0 & \pi-\kappa_{\alpha} & \frac{\gamma_3}{\gamma_0}\pi^* & 0 \\
\pi^*\mp\kappa_{\alpha} & 0 & 0 & \gamma'_1 \\
\frac{\gamma_3}{\gamma_0}\pi & 0 & 0 & \pi^*\pm\kappa_{\alpha} \\
0 & \gamma'_1 & \pi+\kappa_{\alpha} & 0
\end{pmatrix},
\end{equation}
with $\alpha=\parallel$ or $\perp$ indicating the orientation of the in-plane magnetic field, $\pi=k_x-ik_y$, $\kappa_{\parallel}=c/2l_B^2$, $\kappa_{\perp}=ic/2l_B^2$, $l_B=\sqrt{\hbar/eB}$ the magnetic length, $\gamma'_1=\gamma_1/\hbar v_{\rm F}$, $\gamma_3=0.29$ eV and $v_{\rm F}\approx10^6\ \text{m/s}$ the monolayer graphene Fermi velocity.\cite{Neto2009} The upper and lower sign correspond to $\alpha=\parallel$ and $\alpha=\perp$, respectively. The energy spectrum of $H_{\parallel}$ near the K-point is shown in Figs. \ref{fig:energie}(b)-(d) and the energy spectrum of $H_{\perp}$ near the K-point is shown in Figs. \ref{fig:energie}(e)-(f). Their distinctive feature is that the parabolic dispersion describing the electrons in Bernal stacked bilayer graphene\cite{tweeband} is replaced by two monolayer-like Dirac cones shifted along the $k_i$-axis to the positions
\begin{equation}\label{eq:dirac_cone_shift_magnetic_field}
k_i = \pm \kappa,
\end{equation}
with $\kappa=c/2l_B^2$ and with $i=x$ and $i=y$ for $\alpha=\parallel$ and $\alpha=\perp$, respectively. Notice that this shift increases linearly with magnetic field, i.e. $\kappa= 2545\ B[\text{T}]\ \text{cm}^{-1}$.

At low-energy these bands can be described by an approximate $2\times 2$ Hamiltonian\cite{tweeband} which is valid for energies $E\ll \gamma_1$ and is given by
\begin{equation}
\label{tweeband}
H_{2,\alpha} = -\frac{(\hbar v_{\rm F})^2}{\gamma_1}
\begin{pmatrix}
0 & (k_x-ik_y)^2\mp\kappa^2 \\
(k_x+ik_y)^2\mp\kappa^2 & 0
\end{pmatrix},
\end{equation}
where we have neglected $\gamma_3$. The corresponding two-band energy spectrum is shown in Fig. \ref{fig:energie}(c). Eq. (\ref{tweeband}) shows that for $k_y=0$ the Hamiltonian commutes with the Pauli matrix $\sigma_x$, which has eigenvalues $\pm1$. Transforming the above Hamiltonian to the basis formed by the eigenstates $\frac{1}{\sqrt{2}}(1,\pm1)$ of $\sigma_x$ leads to the diagonalised Hamiltonian $H_2=-\frac{(\hbar v_{\rm F})^2}{\gamma_1}(k_x^2\mp\kappa^2)\sigma_z$. Therefore, one can identify the two branches of the low-energy spectrum with the orthonormal pseudospin-up and pseudospin-down eigenstates, respectively. This identification is shown as the red and blue bands in Figs. \ref{fig:energie}(d) and (f). If electrons impinge perpendicularly on a potential step or barrier, the conservation of pseudospin implies that these bands are effectively decoupled and a transition between them is prohibited. Since the effect of in-plane magnetic fields with strengths less than $10^3$ T is only important for energies $E\ll\gamma_1$, we may limit ourselves to the two low-energy bands of the Hamiltonian \eqref{hamiltoniaan}. The low-energy spectrum including the interlayer hopping $\gamma_3$\cite{trigonal,InPlaneField2} is shown in Fig. \ref{fig:energie}(d). This term leads to an additional shift of the Dirac cones, which is of no fundamental importance. The $\gamma_4$ interlayer hopping leads to an asymmetry between electrons and holes. Therefore, the inclusion of these hopping parameters would not significantly affect the physics discussed in this work. Furthermore, the influence of the skew hopping parameters on the transport properties of bilayer graphene was already shown to be negligible\cite{vierband} and because of their relatively small importance but large computational cost they will also be neglected in this work.

\section{Creating pseudo-magnetic fields by shifting one layer}
\label{sec:Creating pseudo-magnetic fields using shifted bilayer graphene}

Fig. (\ref{fig:energie}) shows that in order to have an appreciable effect of the magnetic field on the electronic spectrum, one requires field strengths of several 100's T, which can only be generated by destructive pulsed magnetic fields in a laboratory.\cite{ultrasterk} However, it is possible to obtain a similar electronic spectrum by continuously shifting one of the graphene layers along the armchair direction starting from the Bernal stacking configuration,\cite{shifted} as shown in Fig. \ref{fig:shiftplot}(a). This is equivalent to having a different stacking and a suitable system was recently proposed which can locally be realized experimentally by applying a small strain.\cite{shiftedexp} As shown in Figs. \ref{fig:shiftplot}(b) and (c) this spectrum reproduces correctly the splitting of the parabolic energy band into two linear Dirac cones. The location of these new Dirac cones in reciprocal space is related to the shift $\delta$ as
\begin{equation}
\label{diracshift}
k_ya = \pm\frac{2\gamma_1}{3\gamma_0}\sqrt{1+2\cos\left(\frac{2\pi}{3}\left(1-\frac{\delta}{a_0}\right)\right)}.
\end{equation}
Comparing this relation with the position of the Dirac cones due to the in-plane magnetic field, Eq. (\ref{eq:dirac_cone_shift_magnetic_field}), we can establish a relation between the shift $\delta$ and the strength of the magnetic field $B$ as
\begin{equation}
\label{pseudomagneet}
B = \frac{4\hbar\gamma_1}{3ace\gamma_0}\sqrt{1+2\cos\left(\frac{2\pi}{3}\left(1-\frac{\delta}{a_0}\right)\right)}.
\end{equation}
This relation is shown in Fig. \ref{fig:shiftplot}(d). Because the above formula shows that the low-energy spectrum of a graphene bilayer where one layer is shifted over a distance $\delta$ in the armchair direction corresponds to that of an in-plane magnetic field in the armchair direction, the magnetic field $B$ from Eq. (\ref{pseudomagneet}) can be considered as a \textit{pseudo-magnetic field}. Fig. \ref{fig:shiftplot}(d) shows that the strength of the pseudo-magnetic field can reach values of up to 1500 T. Note that the correspondence of both energy spectra is only valid at low-energy. This is however sufficient for the energy range considered in this study. The energy spectrum for the case of the in-plane magnetic field is simpler and can therefore be used as a theoretical approximation to determine the tunneling properties of shifted bilayer graphene. Vice versa, shifted bilayer graphene can be used as an experimental approximation for bilayer graphene in an in-plane magnetic field.

\section{Spinor wave function}
\label{sec:Spinor wave function}

The scattering potential profiles studied in this paper are the potential step, which models a single gated pn-junction and is given by
\begin{equation}
\label{stap}
V(x) = \left\{
\begin{matrix}
0 & \text{if}\ x<0 & \text{Region 1} \\
VI_4 & \text{if}\ x>0 & \text{Region 2}
\end{matrix}\right.,
\end{equation}
and the potential barrier, which models a single gated pnp-junction and is given by
\begin{equation}
\label{barriere}
V(x) = \left\{
\begin{matrix}
0 & \text{if}\ x<0 & \text{Region 1} \\
VI_4 & \text{if}\ 0<x<d & \text{Region 2} \\
0 & \text{if}\ x>d & \text{Region 3}
\end{matrix}\right.,
\end{equation}
with $V$ the height of the step or barrier, $d$ the width of the barrier and $I_4$ the $4\times4$ identity matrix. Since these potential profiles break translation invariance in the $x$-direction, $k_x$ will no longer be a good quantum number. It is therefore useful to switch to position representation and calculate the wave function from the Schr\"odinger equation $H\Psi=E\Psi$ by solving the set of coupled differential equations. The solution can be written as\cite{multiband}
\begin{equation}
\label{golffunctie}
\Psi(x,y) = \mathcal{P}\mathcal{E}(x,y)\mathcal{C},
\end{equation}
with 
\begin{equation}\label{eq:CMatrix}
\mathcal{E}(x,y)=\text{Diag}[e^{ik_-x},e^{-ik_-x},e^{k_+x},e^{-k_+x}]e^{ik_yy}
\end{equation} 
and where the four component vector $\mathcal{C}$ contains the different coefficients expressing the relative importance of the different exponentials, which have to be set according to the appropriate boundary conditions. The matrix $\mathcal{P}$ from Eq. (\ref{golffunctie}) is
\begin{widetext}
\begin{equation}
\begin{split}
\label{P}
&\mathcal{P} = 
\begin{pmatrix}
(-)^{\zeta}A(-k_-)B(k_-) & (-)^{\zeta}A(k_-)B(-k_-) & (-)^{\zeta}A(ik_+)B(-ik_+) & (-)^{\zeta}A(-ik_+)B(ik_+) \\
B(k_-) & B(-k_-) & B(-ik_+) & B(ik_+) \\
A(k_-) & A(-k_-) & A(-ik_+) & A(ik_+) \\
1 & 1 & 1 & 1
\end{pmatrix}, 
\end{split}
\end{equation}
\end{widetext}
with $\zeta=1$ and $\zeta=0$ for the case of a magnetic field oriented parallel and perpendicular to the potential boundaries, respectively, and where
\begin{eqnarray}
&A(k) = \frac{k+ik_y+\kappa}{\varepsilon}, \\
&B(k) = \frac{\varepsilon^2-k_y^2-(k+\kappa)^2}{\varepsilon\gamma_1'},
\end{eqnarray}
for a magnetic field oriented parallel to the potential boundaries and
\begin{eqnarray}
&A(k) = \frac{k+ik_y-i\kappa}{\varepsilon}, \\
&B(k) = \frac{\varepsilon^2-k^2-(k_y-\kappa)^2}{\varepsilon\gamma_1'},
\end{eqnarray}
for a magnetic field oriented perpendicular to the potential boundaries. The $y$-dependence of the wave function consists of a simple phase factor, which is a consequence of the translational invariance of the system in the $y$-direction and which will not contribute to the transmission probability.

The wave number in the $x$-direction is given by
\begin{equation}
\label{k}
k_{\pm} = \sqrt{\pm(k_y^2-\varepsilon^2-\kappa^2)+\sqrt{4\kappa^2(\varepsilon^2-k_y^2)+\varepsilon^2\gamma_1'^2}},
\end{equation}
for a magnetic field oriented parallel to the potential boundaries and
\begin{equation}
\label{kperp}
k_{\pm} = \sqrt{\pm(k_y^2-\varepsilon^2+\kappa^2)+\sqrt{4\kappa^2k_y^2+\varepsilon^2\gamma_1'^2}},
\end{equation}
for a magnetic field oriented perpendicular to the potential boundaries, with $\varepsilon=E/\hbar v_{\rm F}$. When both $k_-$ and $k_+$ are real numbers the wave function consists of a superposition of a left and a right propagating wave with wave number $k_-$ and two exponential terms with inverse decay length $k_+$, as can be seen from Eq.~(\ref{eq:CMatrix}). However, if $k_-$ becomes imaginary, the corresponding exponential terms will no longer represent propagating waves, while if $k_+$ becomes imaginary the corresponding exponential terms will represent another set of propagating waves. An electron or hole can therefore have 0, 1 or 2 different propagating modes depending on the values of $k_y$, $\varepsilon$ and $\kappa$. 

The wave function inside the potential step or barrier follows from substituting $\varepsilon\rightarrow\varepsilon-\nu$ with $\nu=V/\hbar v_{\rm F}$ in both $\mathcal{P}$ and $k_{\pm}$, which will thus result in a different $x$-component for the wave number, denoted as $q_{\pm}$.

\section{Current density and scattering probabilities}
\label{sec:Current density and scattering probabilities}

\begin{figure}[tb]
\centering
\includegraphics[width=8.5cm]{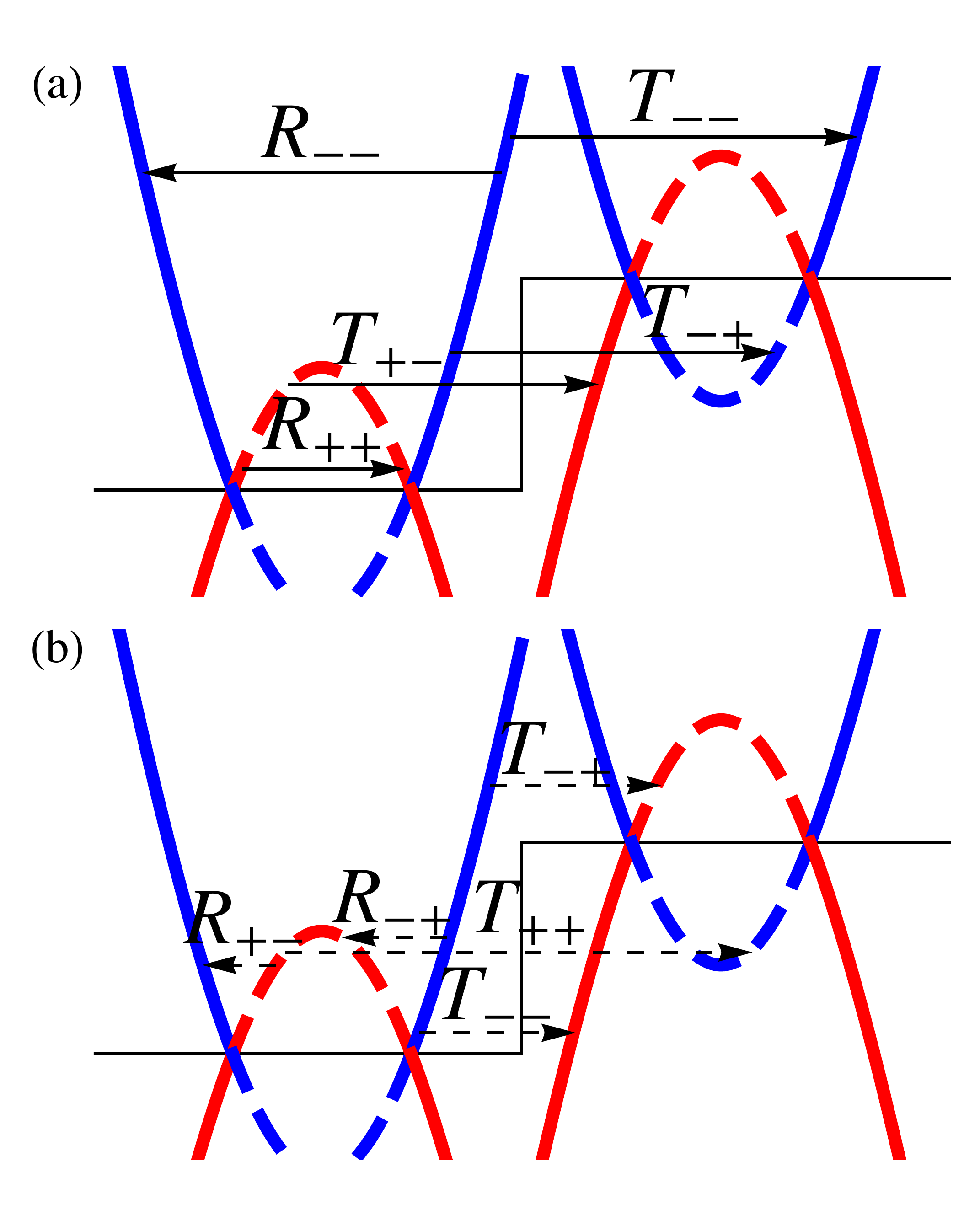}
\caption{(Color online) Schematic representation of the different transmission and reflection probabilities for electrons at perpendicular incidence in the presence of a magnetic field oriented parallel to the potential boundaries. The solid parts of the energy bands are associated with the $k_-$ and $q_-$ propagating modes while the dashed parts of the energy bands are associated with the $k_+$ and $q_+$ propagating modes. The colors of the curves correspond to those of Fig. \ref{fig:energie}(d). (a) Transitions that conserve pseudospin are allowed. (b) Transitions that don't conserve pseudospin are prohibited.}
\label{fig:transschema}
\end{figure}

For a magnetic field oriented parallel to the potential boundaries, at a given energy, different propagating modes exist outside and inside the potential step or barrier, which can be seen from Fig. \ref{fig:energie}(d). We therefore need to consider different transition probabilities between each of them. These 8 different transitions are illustrated in Fig. \ref{fig:transschema}. A transmission to another potential region is denoted by $T_{\xi\eta}$, which represents an electron scattering from the $k_{\xi}$-mode to the $k_{\eta}$-mode, with $\xi,\eta=\pm$. A reflection within the same potential region is denoted by $R_{\xi\eta}$. As a consequence of conservation of pseudospin we have for perpendicular impinging electrons that $T_{--}=T_{++}=0$ for $E<V$, $T_{-+}=T_{+-}=0$ for $E>V$ and $R_{-+}=R_{+-}=0$, as is indicated in Fig. \ref{fig:transschema}.

To find an expression for the different scattering probabilities one first has to determine an expression for the current density of each scattering mode. This expression can be found by inserting the Hamiltonian into the continuity equation. Since the terms involving the in-plane magnetic field form a Hermitian matrix, these will drop out of the equation and the current density is given by 
\begin{equation}
\vec{j}=v_{\rm F}\Psi^{\dag}\vec{\Sigma}\Psi.
\end{equation} 
Here $\Sigma_{x(y)}$ is the $4\times4$ block diagonal matrix with two Pauli matrices $\sigma_{x(y)}$ on the diagonal. Filling in the wave function \eqref{golffunctie} for the general case $\mathcal{C}=(c^{+-},c^{--},c^{-+},c^{++})^T$ in the expression of the $x$-component of the current density results in
\begin{equation}
\label{stroomsom}
j_x = j_x^{+-}+j_x^{--}+j_x^{-+}+j_x^{++},
\end{equation}
where the left superscript refers to a left ($-$) or right ($+$) propagating wave and the right superscript refers to the propagating mode. This means that the total $x$-component of the current density is the sum of the current densities of the separate propagation modes and directions. It should be noted that the current density associated with an exponential term is vanishing. For a magnetic field oriented parallel to the potential boundaries, the first term in Eq. (\ref{stroomsom}) in for example potential region 1, as defined in Eqs. (\ref{stap}) and (\ref{barriere}), is given by
\begin{equation}
\label{stroomvb}
\begin{split}
j^{+-}_{x,1} = &2v_{\rm F}|c_1^{+-}|^2 \\
&\times\Bigg(\frac{\big(\varepsilon^2-k_y^2-(k_-+\kappa)^2\big)^2}{\varepsilon^2\gamma_1'^2}\frac{(k_--\kappa)}{\varepsilon}+\frac{k_-+\kappa}{\varepsilon}\Bigg).
\end{split}
\end{equation}

\begin{figure}[tb]
\centering
\includegraphics[width=8.5cm]{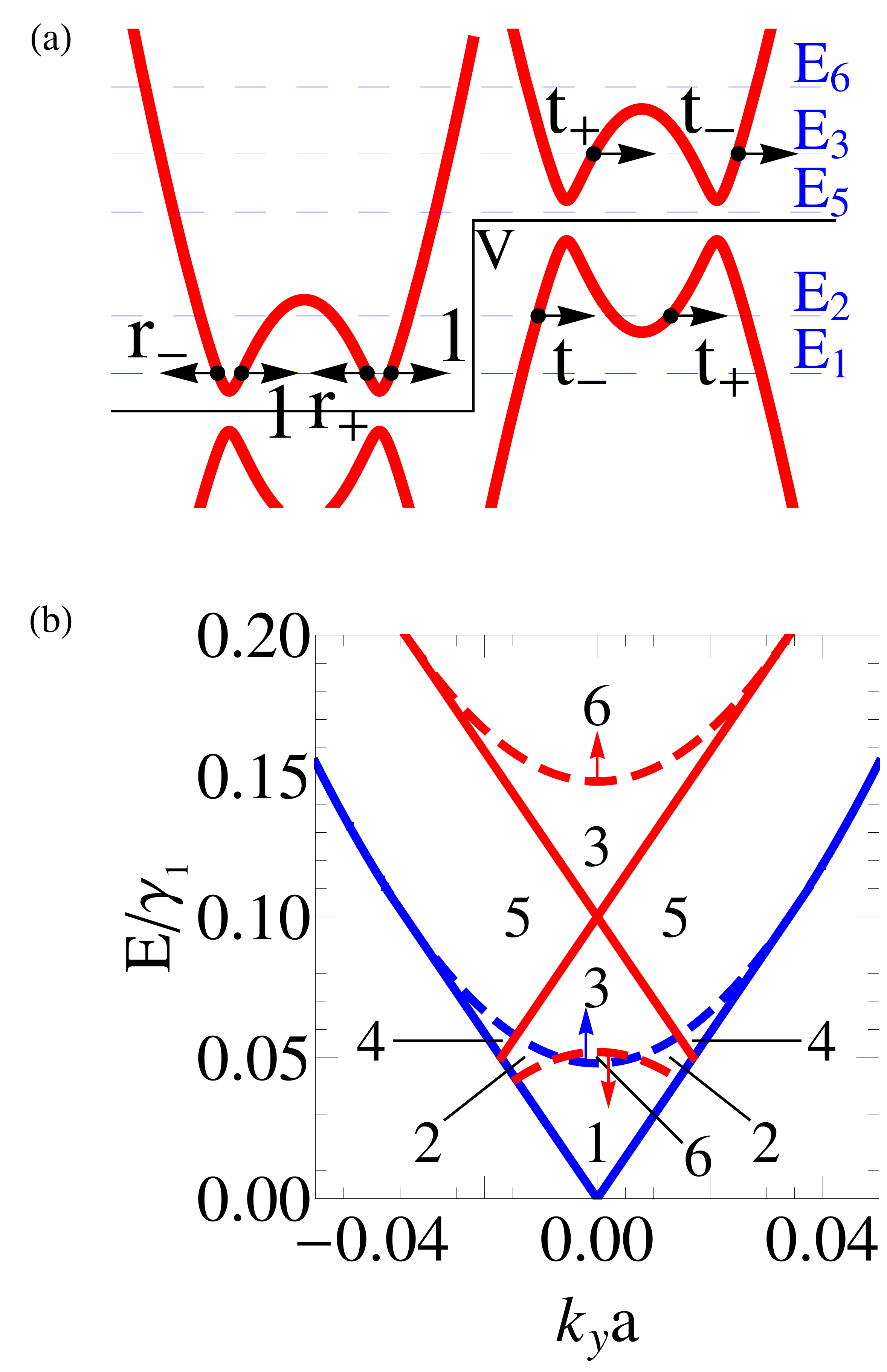}
\caption{(Color online) (a) Schematic representation of electron scattering on a potential step in the presence of a magnetic field oriented parallel to the potential boundaries. The energy spectrum (red) is shown for $k_y\neq0$. The propagating modes (black arrows) and their respective coefficients are shown for different energies, where the subscript of the energies refers to the respective zone in (b) to which the electron corresponds. (b) Schematic representation of the different zones in the $k_y-E$ plane with different propagating modes in the presence of a magnetic field oriented parallel to the potential boundaries. The blue (blue dashed) lines indicate the boundaries for the $k_-$($k_+$)-mode and the red (red dashed) lines indicate the boundaries for the $q_-$($q_+$)-mode. The arrows indicate the direction in which the boundaries will move when the magnetic field becomes larger.}
\label{fig:biparschematisch}
\end{figure}

\begin{figure*}[tb]
\centering
\includegraphics[width=17cm]{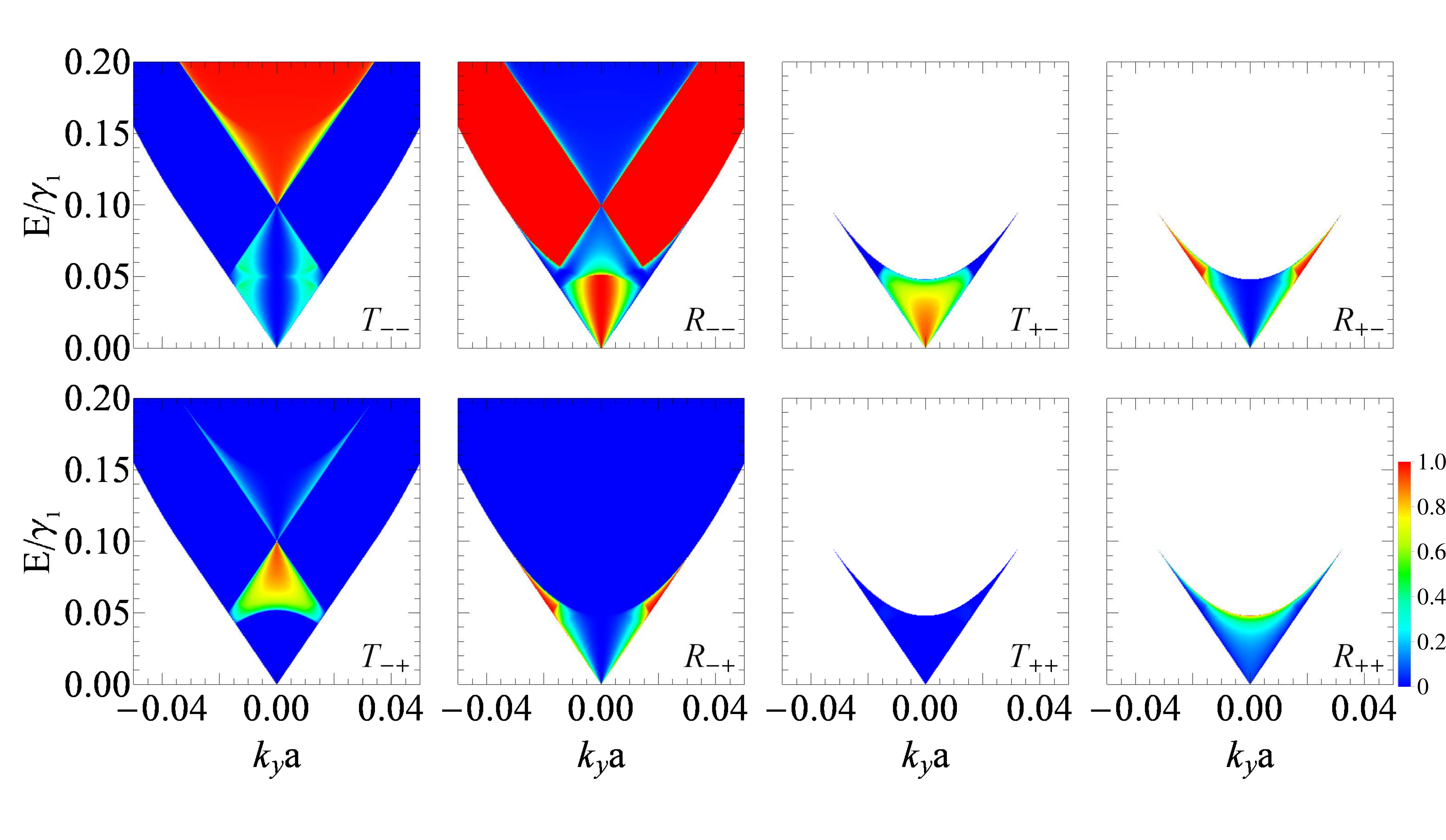}
\caption{(Color online) Transmission and reflection probabilities as a function of the transverse wave number $k_y$ and the energy $E$ in the case of a potential step with height $V=0.1\gamma_1$ in the presence of an in-plane magnetic field oriented parallel to the potential boundaries with strength $B=500$ T.}
\label{fig:grid}
\end{figure*}

\begin{figure*}[tb]
\centering
\includegraphics[width=17cm]{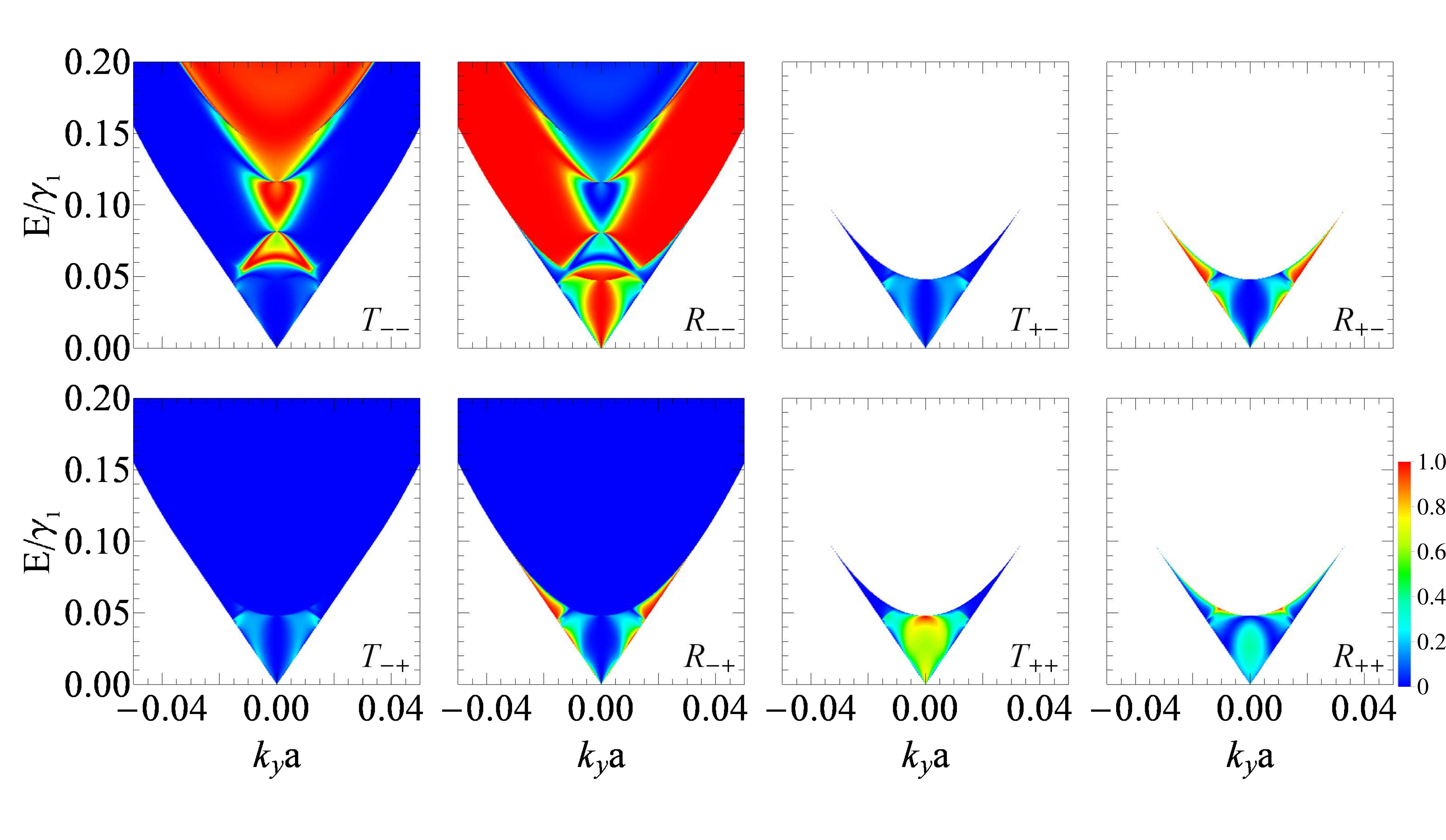}
\caption{(Color online) Transmission and reflection probabilities as a function of the transverse wave number $k_y$ and the energy $E$ in the case of a potential barrier with height $V=0.1\gamma_1$ and width $d=50$ nm in the presence of an in-plane magnetic field oriented parallel to the potential boundaries with strength $B=500$ T.}
\label{fig:gridbar}
\end{figure*}

Since the wave function is a stationary state and the $y$-dependence consists of a simple phase factor meaning that $\vec{j}$ does not depend on $y$, the continuity equation implies that the current density $j_{x}$ is constant in the $x$-direction. Expressing this conservation for the two potential regions of the potential step and dividing both sides by $j_x^{+-}$ yields
\begin{equation}
\label{stroombehoud}
\frac{j^{+-}_{x,2}}{j^{+-}_{x,1}}+\frac{j^{--}_{x,2}}{j^{+-}_{x,1}}+\frac{j^{-+}_{x,2}}{j^{+-}_{x,1}}+\frac{j^{++}_{x,2}}{j^{+-}_{x,1}}-\frac{j^{--}_{x,1}}{j^{+-}_{x,1}}-\frac{j^{-+}_{x,1}}{j^{+-}_{x,1}}-\frac{j^{++}_{x,1}}{j^{+-}_{x,1}} = 1.
\end{equation}
The same expression, but with the subscript 3 instead of 2, follows when equating the $x$-component of the current density of potential region 1 and 3 of the potential barrier. 

When considering physical scattering processes, however, appropriate boundary conditions should be imposed. It is useful to consider an incoming electron in a single propagation mode, for example the $k_-$-mode. The $c^{+-}$ coefficient then corresponds to the incoming electron and can be normalised to 1. Since we assume that there is no other incoming electron, the $c^{++}$ coefficient has to be set to 0. When the $k_+$-mode isn't propagating, the corresponding term is an exponential which diverges for $x\rightarrow-\infty$ and thus also has to be excluded from the calculation by again setting $c^{++}=0$. The other two terms represent either a reflected wave or an exponential term that doesn't diverge and their coefficients will be denoted as the reflection coefficients $r_-$ and $r_+$. For potential region 1 one can therefore write $\mathcal{C}_1=(1,r_-,r_+,0)^T$. When considering an incoming electron in the $k_+$-mode the boundary condition is $\mathcal{C}_1=(0,r_-,r_+,1)^T$.

At the right side of the potential step (region 2) for $E>V$ or barrier (region 3) there can be no left propagating wave and therefore one has to set both $c^{--}$ and $c^{-+}$ to 0. When $q_+$ (step) or $k_+$ (barrier) is not a propagating mode, the coefficient $c^{-+}$ corresponds to an exponential term that diverges for $x\rightarrow\infty$ and therefore again has to be set to 0. The other two terms represent either a transmitted wave or an exponential term that doesn't diverge and their coefficients will be denoted as the transmission coefficients $t_-$ and $t_+$. One can therefore write $\mathcal{C}_{2(3)}=(t_-,0,0,t_+)^T$ for the potential step (barrier). For $E<V$ this becomes $\mathcal{C}_2=(0,t_-,t_+,0)^T$ for the potential step when the $q_+$-mode is propagating and $\mathcal{C}_2=(0,t_-,0,t_+)^T$ when it isn't. Inside the potential barrier all the different terms (propagating or not) can be present and therefore one has in general $\mathcal{C}_2=(c_2^{+-},c_2^{--},c_2^{-+},c_2^{++})^T$.

The different propagating modes along with their coefficients are indicated in Fig. \ref{fig:biparschematisch}(a). With these boundary conditions the second, third and last term of the left hand side of Eq. \eqref{stroombehoud} vanish and it can therefore be rewritten as $T_{--}+T_{-+}+R_{--}+R_{-+}=1$. This expresses a specific form of the conservation of probability
\begin{equation}
\label{normalisatie}
\sum_{\eta=\pm}\big(T_{\xi\eta}+R_{\xi\eta}\big) = 1,
\end{equation}
where the different scattering probabilities are defined as
\begin{equation}
\label{scattering}
T_{\xi\eta} = \frac{j_{x,2}^{+,\eta}}{j_{x,1}^{+,\xi}}, \quad\quad 
R_{\xi\eta} = \frac{|j_{x,1}^{-,\eta}|}{j_{x,1}^{+,\xi}},
\end{equation}
with $\xi,\eta=\pm$.

For a magnetic field oriented perpendicular to the potential boundaries the situation is much simpler since at a given energy $E<\gamma_1$ maximum two propagating modes exist outside and inside the potential step or barrier, which can be seen from Fig. \ref{fig:energie}(f). This means that for energies $E<\gamma_1$ the $k_+$-mode is always evanescent and therefore the transport properties are determined by a single transmission probability $T\equiv T_{--}$ and a single reflection probability $R\equiv R_{--}$. Furthermore, since conservation of probability requires that $T+R=1$, the reflection probability follows trivially once the transmission probability is known. Therefore, we can restrict ourselves to the transmission probability. In this case the current density is given by
\begin{equation}
\label{stroomvbperp}
j^{+-}_{x,1} = 2v_{\rm F}|c_1^{+-}|^2\frac{k_-}{\varepsilon}\Bigg(\frac{\big(\varepsilon^2-k_-^2-(k_y-\kappa)^2\big)^2}{\varepsilon^2\gamma_1'^2}+1\Bigg).
\end{equation}
The boundary conditions for this situation are $\mathcal{C}_1=(1,r_,c,0)^T$, $\mathcal{C}_{2(3)}=(t,0,0,d)^T$ for the potential step for $E>V$ or barrier, $\mathcal{C}_2=(0,t,0,d)^T$ for the potential step for $E<V$ and $\mathcal{C}_2=(c_2^{+-},c_2^{--},c_2^{-+},c_2^{++})^T$ for the potential barrier.

Using these definitions the different scattering probabilities can be calculated from the reflection and transmission coefficients, which can in turn be determined by equating the wave functions from the regions outside and inside the potential step or barrier at the boundaries $x=0$ and $x=d$.

\section{Numerical results}
\label{sec:Numerical results}

\subsection{Scattering probabilities}
\label{subsec:Scattering probabilities}

\begin{figure*}[tb]
\centering
\includegraphics[width=17cm]{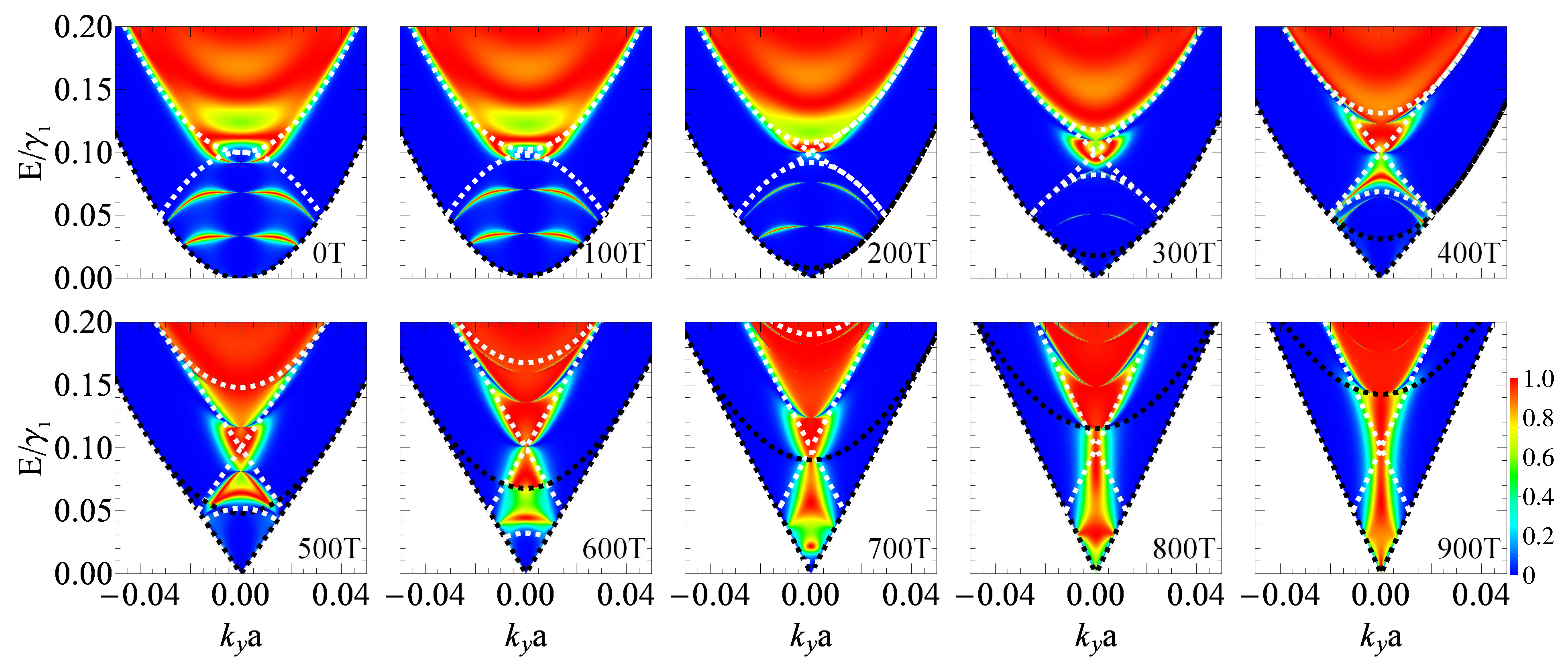}
\caption{(Color online) Transmission probability $T_{--}$ as a function of the transverse wave number $k_y$ and the energy $E$ in the case of a potential barrier with height $V=0.1\gamma_1$ and width $d=50$ nm in the presence of an in-plane magnetic field oriented parallel to the potential boundaries with strength varying from $B=0$ T to $B=900$ T as indicated in the different panels. The black dashed curves indicate the boundaries for the $k_-$- and $k_+$-modes and the white dashed curves indicate the boundaries for the $q_-$- and $q_+$-modes.}
\label{fig:Tmmevolutie}
\end{figure*}

The results for the different transmission and reflection probabilities for electron scattering on a potential step with the application of an in-plane magnetic field oriented parallel to the potential boundaries are shown in Fig. \ref{fig:grid}. The probabilities are shown as a function of the conserved quantities $k_y$ and $E$. If a part of the $k_y-E$ plane in Fig. \ref{fig:grid} is coloured in white this means that the incoming mode of the respective scattering probability is not propagating and therefore the probability is undefined. The results of Fig. \ref{fig:grid} can be understood by means of Fig. \ref{fig:biparschematisch}(b), in which the $k_y-E$ plane is divided in different zones that are defined by different propagating modes as we will explain in this section. 

In zone 1, the modes corresponding to $k_-$, $k_+$ and $q_-$ are propagating while the $q_+$-mode is not. This explains why $T_{-+}$ and $T_{++}$ are vanishing in that zone while the other probabilities are finite. Indeed, the probabilities $T_{-+}$ and $T_{++}$ correspond to modes moving inwards with wave number $k_-$ or $k_+$, respectively, and scatter to the $q_+$-mode afterwards. Because the latter mode is not propagating, the transmission probabilities will vanish. 

Transiting to zone 2 the $q_+$-mode also becomes propagating and therefore $T_{-+}$ is non vanishing. As a consequence of conservation of probability, Eq. \eqref{normalisatie}, this lowers the other scattering probabilities.

In zone 3 the $k_+$-mode is not propagating and therefore in this zone the scattering probabilities that correspond to an electron incoming in the $k_+$-mode are not defined. This also explains why $R_{-+}=0$ in this zone. 

In zone 4 all the modes outside the potential are propagating while all the modes inside the potential are not. Therefore, all the transmission probabilities vanish while the four reflection probabilities do not. Note that in this zone the electrons rather scatter into the other mode upon reflection. 

The only propagating mode in zone 5 is the $k_-$-mode which is why $R_{--}=1$ in this zone while all the other scattering probabilities either vanish or are undefined.

Finally, in zone 6 the $k_-$- and $q_-$-modes are propagating while their $k_+$- and $q_+$-counterparts are not. This explains why $T_{-+}$ and $R_{-+}$ vanish while $T_{--}$ and $R_{--}$ are finite. However, because the energy of the electrons is much larger than the height of the potential step, the transmission is nearly unity while the reflection is almost zero.

Fig. \ref{fig:grid} shows that $R_{-+}=R_{+-}$ where both probabilities are defined. This is a consequence of time reversal symmetry and the fact that the two different valleys in the graphene Brillouin zone are equivalent.\cite{vierband} Furthermore, it is clear that the results of the scattering probabilities at normal incidence are in agreement with the predictions made in Sec. \ref{sec:Current density and scattering probabilities} based on the conservation of pseudospin.

The results for electrons scattering on a potential barrier are shown in Fig. \ref{fig:gridbar}. The reflection probabilities are similar to those for the case of scattering on a potential step since the situation at the left side of the potential has not changed. In contrast, the transmission probabilities differ significantly from those for the potential step. 

Fig. \ref{fig:gridbar} shows that the transmission probabilities for which the propagating mode is the same at the left and right side of the barrier are dominant over the transmission probabilities for which the propagating mode is altered. For example, while the $T_{+-}$ channel for low-energy is significant for the potential step, for the barrier it is almost completely suppressed to the advantage of the $T_{++}$ channel in the same energy range. A similar probability transfer has occurred between the $T_{--}$ and $T_{-+}$ channels in the lower part of zone 3. The reason for this probability transfer is that a possible way of transiting the barrier region in the same mode is by switching modes when entering and leaving the barrier, i.e. $k_{\pm}\rightarrow q_{\mp}\rightarrow k_{\pm}$. Fig. \ref{fig:grid} shows that in the case of the potential step, the transmission probabilities $T_{\pm\mp}$ dominate over the probabilities $T_{\pm\pm}$ for $E<V$. Thus in the case of the potential barrier, for $E<V$ the transmission probabilities $T_{\pm\pm}$ will dominate because the electron is likely to change its propagating mode at both edges of the potential. The reason that $T_{--}$ is dominant for $E>V$ in the case of the potential barrier is because $T_{--}$ is also dominant for $E>V$ in the case of the potential step and thus a $k_-$ electron can go through the barrier by staying in the same mode. 

Furthermore, it is clear from Fig. \ref{fig:gridbar} that now not only $R_{-+}=R_{+-}$, but also $T_{-+}=T_{+-}$. The reason is similar to the argument used for the case of the potential step.\cite{vierband} 

A remarkable feature of the transmission probability $T_{--}$ is that in a range around $E=V$ and $k_y=0$ there is a non vanishing transmission probability despite the fact that there are no propagating modes inside the barrier. This finite transmission stems from evanescent tunneling through the finite barrier, a process which is impossible for the step since it is infinitely wide. 

Furthermore, the results are again in agreement with the predictions made in Sec. \ref{sec:Current density and scattering probabilities} based on the conservation of pseudospin, keeping in mind that the predictions for $E<V$ are not valid for the barrier and that the predictions for $E>V$ can be extended to the entire positive energy domain since the electron will always end up in a positive energy solution at the right side of the barrier. This is in contrast to the case of the potential step for which the electron will end up in a negative energy solution in the step for $E<V$.

Fig. \ref{fig:gridbar} shows that the transmission probability $T_{--}$ carries most of the tunneling features present in the system. Therefore, in Fig. \ref{fig:Tmmevolutie} we show the evolution of this quantity with increasing magnetic field. On top of the numerical results we have plotted the curves from Fig. \ref{fig:biparschematisch} that define the different zones. The results show that as the in-plane magnetic field increases, the typical bilayer graphene resonances at $E<V$ shift upwards in energy, become sharper and eventually vanish at about $300$ T. As the magnetic field increases further the chiral suppressed tunneling at normal incidence that is characteristic for bilayer graphene is replaced by strong chiral supported tunneling, typical for monolayer graphene. This happens in the region corresponding to zone 2 in Fig. \ref{fig:biparschematisch}(b) and is to be expected as in this zone the energy spectrum can both inside and outside the potential barrier be considered to be that of two separate linear Dirac cones. The requirement that the black and white dashed lines in Fig. \ref{fig:Tmmevolutie} have the same value for $k_y=0$ immediately yields the minimal magnetic field at which chiral supported tunneling for $E<V$ can be observed in this system, which is given by
\begin{equation}
\label{minmag}
B_{\text{min}} = \frac{2\hbar}{ace}\frac{\sqrt{V^2+2\gamma_1V}}{\sqrt{3}\gamma_0},
\end{equation}
which yields $B_{\text{min}}=510.73$ T for $V=0.1\gamma_1$. Note however that at these large magnetic fields also the high energy transmission is affected. From $600$ T and beyond there is a set of anti-resonances appearing in zone 3. These resonances cut into the high transmission region and can be considered as a suppression of transmission due to bound states in the barrier.

\begin{figure*}[tb]
\centering
\includegraphics[width=17cm]{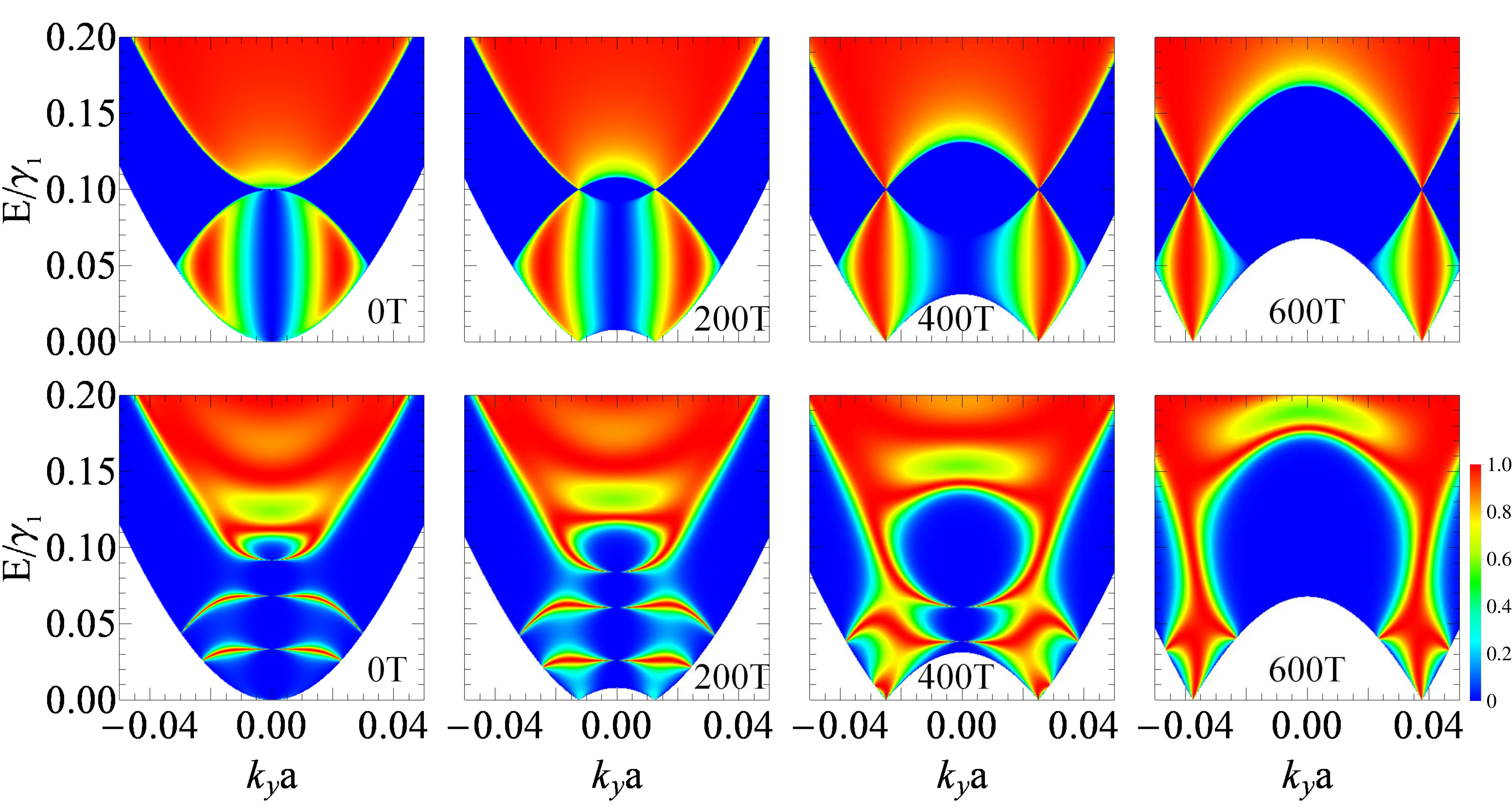}
\caption{(Color online) Top row: Transmission probability as a function of the transverse wave number $k_y$ and the energy $E$ in the case of a potential step with height $V=0.1\gamma_1$ in the presence of an in-plane magnetic field oriented perpendicular to the potential boundaries with strength varying from $B=0$ T to $B=600$ T as indicated in the different panels. Bottom row: The same as the top row but now in the case of a potential barrier with height $V=0.1\gamma_1$ and width $d=50$ nm.}
\label{fig:gridpctot}
\end{figure*}

\begin{figure}[tb]
\centering
\includegraphics[width=8.5cm]{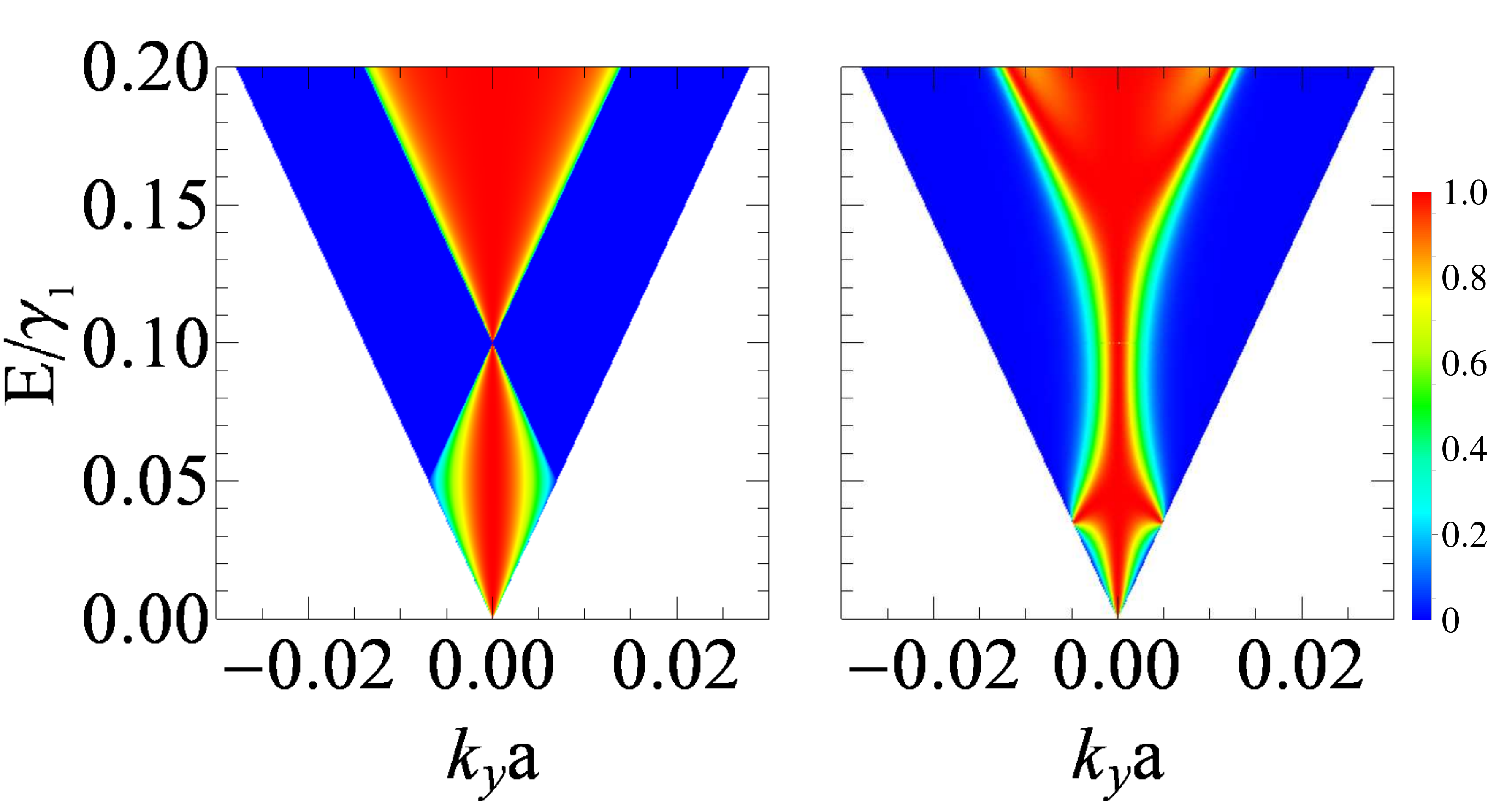}
\caption{(Color online) (a) Transmission probability for monolayer graphene as a function of the transverse wave number $k_y$ and the energy $E$ in the case of a potential step with height $V=0.1\gamma_1$. (b) The same as (a) but now in the case of a potential barrier with height $V=0.1\gamma_1$ and width $d=100$ nm.}
\label{fig:plotmonogrid}
\end{figure}

For the case of a magnetic field oriented perpendicular to the potential boundaries the results are shown in Fig. \ref{fig:gridpctot} for electron scattering on a potential step and barrier. For both cases, one can see from the figure that the transport properties change from a bilayer system, with the typical anti-Klein tunneling for $E<V$ at $k_y=0$ and Klein tunneling for $E<V$ at finite $k_y$,\cite{multilaag} to two separated monolayer-like systems as the magnetic field increases, with Klein tunneling emerging at $k_y=\pm\kappa$ for large enough magnetic fields. This is because, as seen in Fig. \ref{fig:energie}(f), the low-energy spectrum becomes cone-like for $k_y=\pm\kappa$. Moreover, as $k_y$ is conserved because of translational invariance this means that in this case the charge carriers will scatter from cone to cone. As a reference, the transmission probability for electrons in monolayer graphene is shown in Fig. \ref{fig:plotmonogrid} for scattering on a potential step and on a potential barrier.\cite{klein} The reason that the monolayer-like transmission probability for large magnetic fields is similar to that of electrons in monolayer graphene scattering through a potential barrier with width $100$ nm, as opposed to $50$ nm, is because of the reduced Fermi velocity of the split Dirac cones in the presence of an in-plane magnetic field. For the case of the potential step, Fig. \ref{fig:gridpctot} shows that the two high-transmission regions for $E<V$, which are typical for bilayer graphene, shift away from each other and each form a Klein tunneling region at $k_y=\pm\kappa$ of the separated monolayer systems. For the potential barrier the bilayer graphene resonances for $E<V$ broaden and merge with each other to form the monolayer resonances and Klein tunneling at $k_y=\pm\kappa$. Furthermore, the pseudospin structure shown in Fig. \ref{fig:energie}(f) implies that $T=0$ for $E<V$ at normal incidence, which is confirmed by our numerical results.

\subsection{Conductance}
\label{subsec:Conductance}

\begin{figure}[tb]
\centering
\includegraphics[width=8.5cm]{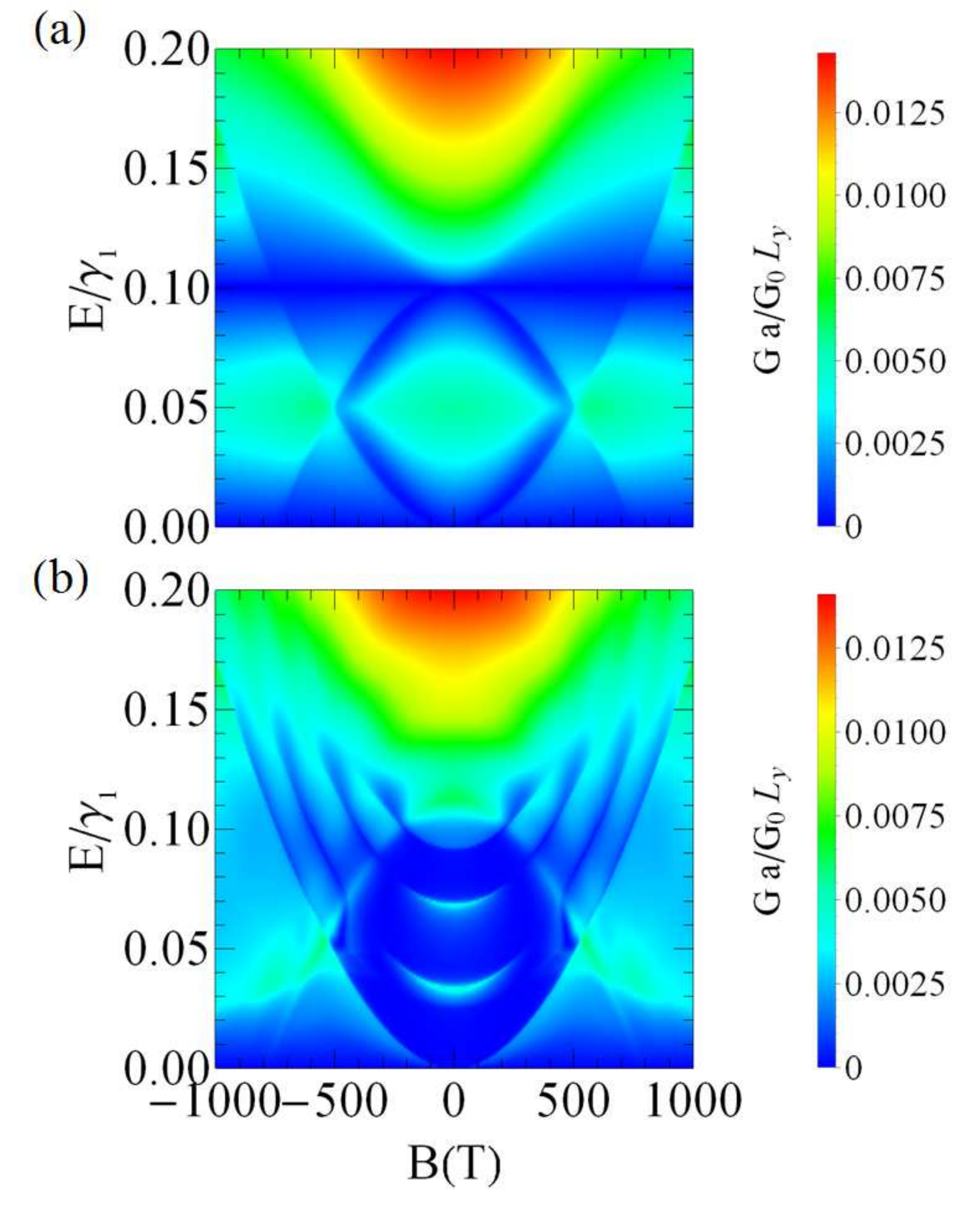}
\caption{(Color online) (a) Contour plot of the conductance as a function of the energy of the electron and the in-plane magnetic field oriented parallel to the potential boundaries in the case of a potential step with height $V=0.1\gamma_1$. (b) Same as (a) but now in the case of a potential barrier with height $V=0.1\gamma_1$ and width $d = 50$ nm.}
\label{fig:conductantiecontour}
\end{figure}

\begin{figure}[tb]
\centering
\includegraphics[width=8.5cm]{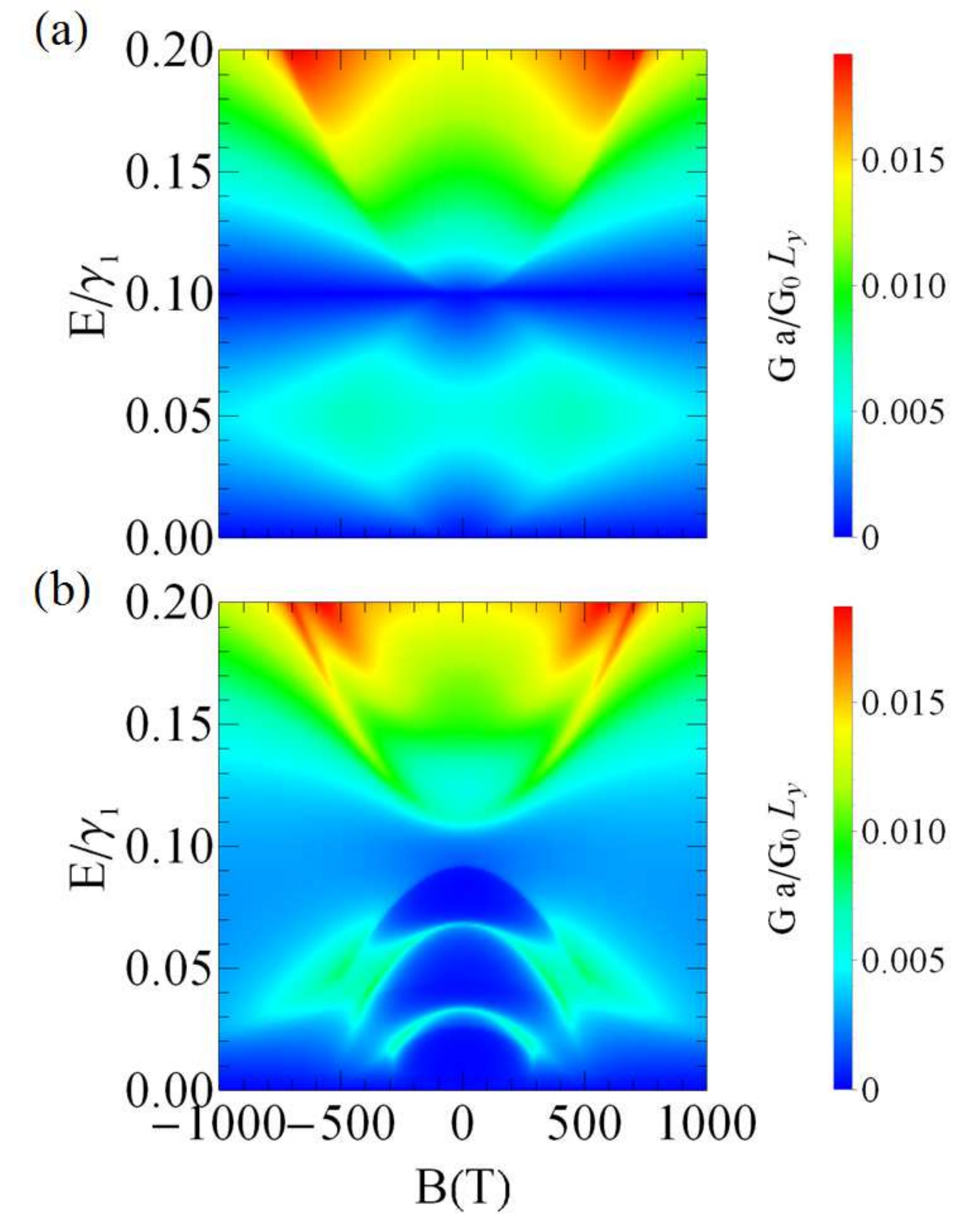}
\caption{(Color online) (a) Contour plot of the conductance as a function of the energy of the electron and the in-plane magnetic field oriented perpendicular to the potential boundaries in the case of a potential step with height $V=0.1\gamma_1$. (b) Same as (a) but now in the case of a potential barrier with height $V=0.1\gamma_1$ and width $d = 50$ nm.}
\label{fig:conductantiecontourpc}
\end{figure}

Using the numerical results for the transmission probabilities, we can obtain the zero temperature conductance from the Landauer-B\"uttiker formula\cite{conductantie, Snyman2007}
\begin{equation}
\label{conductantie}
\frac{G}{G_0} = \frac{L_y}{2\pi}\int dk_y\sum_{\xi,\eta=\pm}T_{\xi\eta},
\end{equation}
where $G_0=4e^2/h$ is four times the quantum of conductance due to spin and valley degeneracy and $L_y$ is the width of the sample in the $y$-direction. The result is shown in Figs. \ref{fig:conductantiecontour}(a) and (b) for the case of a potential step and barrier, respectively, and for a magnetic field oriented parallel to the potential boundaries and in Figs. \ref{fig:conductantiecontourpc}(a) and (b) for the case of a potential step and barrier, respectively, and for a magnetic field oriented perpendicular to the potential boundaries. In these figures we relate three directly measurable quantities, namely the conductance, magnetic field and Fermi energy for an electron tunneling through a pn-junction and a pnp-junction. 

For a magnetic field oriented parallel to the potential boundaries, local maxima and minima for energy values symmetric on either side of $E=V/2$ are present for magnetic fields smaller than $B=500$ T for the potential step. These extrema are a consequence of the fact that the $k_+$-mode ($q_+$-mode) becomes propagating at energies lower (higher) than the minimum below (above) $E=V/2$, as indicated by the $k_+$ ($q_+$) zone boundary in Fig. \ref{fig:biparschematisch}(b). The stronger the magnetic field, the closer these extrema are located to $E=V/2$, in correspondence to the movement of the zone boundaries as indicated by the arrows in Fig. \ref{fig:biparschematisch}(b). At $B=500$ T the extrema have reached $E=V/2$, corresponding to the $k_+$ and $q_+$ zone boundaries passing each other in $k_y=0$ in Fig. \ref{fig:biparschematisch}(b), and there is a region around $E=V/2$ in which both $k_+$ and $q_+$ correspond to propagating modes and as a consequence there is a higher conductance. This region corresponds to zone 2 in Fig. \ref{fig:biparschematisch}(b) and expands as the magnetic field increases. For a constant $E>V$ it is clear from the figure that the conductance decreases with increasing magnetic field. This is a consequence of the fact that as the magnetic field increases, for a given energy there is a smaller $k_y$-interval in which the $q_-$-mode is propagating and thus a smaller $k_y$-interval in which there is a high transmission probability.

For the potential barrier the conductance has different characteristics as a function of the magnetic field for energies $E<V$. The typical bilayer resonances are shifted to higher energy values until they vanish and eventually morph into anti-resonances. This is because the resonances in the transmission probability shift to higher energies as the magnetic field increases, as shown in Fig. \ref{fig:Tmmevolutie}, whilst also becoming narrower and, therefore, contributing less to the conductance. As a consequence, the conductance is very low for $E<V$ in the region where both $k_+$ and $q_+$ are evanescent modes and where the resonances have disappeared. However, when the $k_+$- and $q_+$-modes become propagating, this leads to a higher conductance. In particular, the enhanced conductance in the region corresponding to zone 2 in Fig. \ref{fig:biparschematisch}(b) is again clearly visible. The conductance of electrons with energy $E>V$, similar to the case of the potential step, decreases as the magnetic field becomes larger which is a consequence of the $k_-$-mode being a propagating mode in a decreasing $k_y$-interval. Finally, the anti-resonances in the conductance are the consequence of anti-resonances in the transmission probability as discussed before. The conductance calculation shows the remarkable fact that the resonances in the conductance morph into anti-resonances, implying that they stem from the same physical process and that the resonances in the transmission probability morph into anti-resonances as well. These results are characteristic experimental features and are a direct consequence of the in-plane (pseudo-)magnetic field.

When the thickness of the barrier increases, more resonances and anti-resonances will appear. In the limit of an infinitely thick potential barrier, i.e. a potential step, all the resonances will merge to form the region of enhanced conductance around $E=V/2$ and all the anti-resonances will merge to form the region of zero conductance around $E=V$.

For a magnetic field oriented perpendicular to the potential boundaries, the conductance at first increases as the magnetic field increases for a given energy but then decreases for even larger magnetic fields. The initial increase in the conductance is due to the splitting of the bilayer system into two monolayer-like systems. However, as the magnetic field increases even further the two monolayer-like transmission regions separate further as well. As the transmission probability for $k_y$ values in between these two regions vanishes this leads to a decrease in the conductance. Furthermore, for the case of the potential barrier the resonances for $E<V$ now move to lower energies as the magnetic field increases, as opposed to the case of a magnetic field oriented parallel to the potential boundaries where the resonances move to higher energies before morphing into anti-resonances. There are now also distinct resonances visible for $E>V$.

\section{Summary and conclusion}
\label{sec:Conclusion}

By shifting one of the two layers of the bilayer system, the energy spectrum becomes similar to the spectrum of bilayer graphene in the presence of an in-plane magnetic field. Therefore the shift has the same effect as an in-plane pseudo-magnetic field that can reach extreme high values.

The scattering of electrons on a potential step and barrier in bilayer graphene was studied in the presence of such large in-plane magnetic fields oriented both parallel and perpendicular to the potential boundaries and it was linked with the concept of pseudospin.

For a magnetic field oriented parallel to the potential boundaries, the different scattering probabilities were shown to be drastically altered by the presence of the magnetic field as a consequence of an extra propagating mode at low energies. The results for electrons at normal incidence were explained by conservation of pseudospin and the results in general can be understood by dividing the $k_y-E$ plane in different zones in which electronic transport is governed by different propagating modes. At large magnetic fields, we have further shown that due to the splitting of the parabolic cone into two linear ones, the chiral suppressed transmission is transformed into a chiral supported transmission at normal incidence.

For both a potential step and a potential barrier the conductance decreases as a function of the strength of the parallel magnetic field for energies $E>V$. For electrons with energy $E<V$ the conductance is enhanced in the regions where additional propagating modes are available. Furthermore, in the case of a potential barrier a parallel magnetic field shifts the tunneling resonances to higher energies and eventually morphs them into anti-resonances. This can be clearly seen in the conductance of the system.

For a magnetic field oriented perpendicular to the potential boundaries, we found a very clear transition from a bilayer system to two separated monolayer-like systems. This transition is clearly shown in the shifting of the bilayer graphene high-transmission regions at $E<V$ which leads to the occurrence of Klein tunneling at $k_y=\pm\kappa$ for the case of the potential step. For the potential barrier the bilayer graphene resonances broaden and merge to form monolayer-like resonances and again Klein tunneling at $k_y=\pm\kappa$. These features also leave a signature in the conductance of the system, which should be observable experimentally.

\section{Acknowledgments}

This work was supported by Fonds Wetenschappelijk Onderzoek (FWO-Vl) through an aspirant research grant to MVDD and BVD.


\begin{thebibliography}{10}

\bibitem{ontdekking}
K. S. Novoselov, A. K. Geim, S. V. Morozov, D. Jian, Y. Zhang, S. V. Dubonos, I. V. Grigorieva, and A. A. Firsov, Science \textbf{306}, 666 (2004).

\bibitem{klein}
M. I. Katsnelson, K. S. Novoselov, and A. K. Geim, Nat. Phys. \textbf{2}, 620 (2006).

\bibitem{kleinexp}
N. Stander, B. Huard, and D. Goldhaber-Gordon, Phys. Rev. Lett. \textbf{102}, 026807 (2009).

\bibitem{kleinexp2}
A. F. Young and P. Kim, Nat. Phys. \textbf{5}, 222 (2009).

\bibitem{vierband}
B. Van Duppen and F. M. Peeters, Phys. Rev. B \textbf{87}, 205427 (2013).

\bibitem{Tudorovskiy2012}
T. Tudorovskiy, K. J. A. Reijnders, and M.I. Katsnelson, Phys. Scr. {\bf T146,} 014010 (2012).

\bibitem{Snyman2007} I. Snyman and C. W. J. Beenakker, Phys. Rev. B {\bf 75}, 045322 (2007).

\bibitem{multiband}
B. Van Duppen, S. H. R. Sena, and F. M. Peeters, Phys. Rev. B \textbf{87}, 195439 (2013).

\bibitem{multilaag}
B. Van Duppen and F. M. Peeters, Europhys. Lett. \textbf{102}, 27001 (2013).

\bibitem{pseudo}
W.-Y. He, Y. Su, M. Yang, and L. He, Phys. Rev. B \textbf{89}, 125418 (2014).

\bibitem{InPlaneField2} B. Roy and K. Yang, Phys. Rev. B {\bf 88}, 241107(R) (2013).

\bibitem{monoloodrecht}
M. Ramezani Masir, P. Vasilopoulos, and F. M. Peeters, Phys. Rev. B \textbf{82}, 115417 (2010).

\bibitem{biloodrecht}
M. Ramezani Masir, P. Vasilopoulos, and F. M. Peeters, Phys. Rev. B \textbf{79}, 035409 (2009).

\bibitem{InPlaneField1} S. S. Pershoguba and V.M.Yakovenko, Phys. Rev. B {\bf 82}, 205408 (2010).

\bibitem{condexp}
N. M. R. Peres, Rev. Mod. Phys. \textbf{82}, 2673 (2010).

\bibitem{condexp2}
J. Ye, M. F. Craciun, M. Koshino, S. Russo, S. Inoue, H. Yuan, H. Shimotani, A. F. Morpurgo, and Y. Iwasa, Proc. Natl. Acad. Sci. USA \textbf{108}, 13002 (2011).

\bibitem{Guinea2010}
F. Guinea, M. I. Katsnelson, and A. K. Geim, Nat. Phys. {\bf 6}, 30 (2010).

\bibitem{zelfde2}
A. Daboussi, L. Mandhour, J. N. Fuchs, and S. Jaziri, Phys. Rev. B \textbf{89}, 085426 (2014).

\bibitem{bernal}
J. D. Bernal, Proc. R. Soc. A, \textbf{106}, 749 (1924).

\bibitem{parameters}
B. Partoens and F. M. Peeters, Phys. Rev. B \textbf{74}, 075404 (2006).

\bibitem{tightbinding}
E. McCann and M. Koshino, Rep. Prog. Phys. \textbf{76}, 056503, (2013).

\bibitem{Neto2009}
A. H. Castro Neto, N. M. R. Peres, K.S. Novoselov, and A. K. Geim, Rev. Mod. Phys. {\bf 81}, 109 (2009).

\bibitem{tweeband}
E. McCann and V. I. Fal'ko, Phys. Rev. Lett. \textbf{96}, 086805 (2006).

\bibitem{trigonal}
E. McCann, D. S. L. Abergel, and V. I. Fal'ko, Solid State Commun. \textbf{143}, 110 (2007).

\bibitem{ultrasterk}
R. J. Nicholas, P. Y. Solane, and O. Portugall, Phys. Rev. Lett. \textbf{111}, 096802 (2013).

\bibitem{shifted}
M. Koshino, Phys. Rev. B \textbf{88}, 115409 (2013).

\bibitem{shiftedexp}
J. Lin, W. Fang, W. Zhou, A. R. Lupini, J. C. Idrobo, J. Kong, S. J.
Pennycook, and S. T. Pantelides, Nano Lett. \textbf{13}, 3262 (2013).

\bibitem{conductantie}
Y. M. Blanter and M. B\"uttiker, Phys. Rep. \textbf{336}, 1 (2000).

\end{thebibliography}
\end{document}